\documentclass[pss,fleqn]{w-art}
\usepackage{times}
\usepackage{w-thm}
\usepackage[]{graphicx}
\begin{document}
\DOIsuffix{theDOIsuffix}
\Volume{XX} \Issue{1} \Copyrightissue{01} \Month{01} \Year{2004}
\pagespan{1}{}
\Receiveddate{\sf zzz} \Reviseddate{\sf zzz} \Accepteddate{\sf
zzz} \Dateposted{\sf zzz}


\title[Manuscript preparation guidelines]{Symmetry breaking and  
Wigner molecules in few-electron quantum dots}


\author[Constantine Yannouleas]{Constantine Yannouleas\footnote{e-mail: 
{\sf Constantine.Yannouleas@physics.gatech.edu}}}
\address[]{School of Physics, Georgia Institute of Technology,
Atlanta, GA 30332-0430}
\author[Uzi Landman]{Uzi Landman\footnote{e-mail: 
{\sf Uzi.Landman@physics.gatech.edu}}}


\begin{abstract}
\begin{center}
{\bf [Invited talk at TNT2005 (Trends in Nanotechnology), Physica status solidi (a) 203, 1160 (2006)]}
\end{center}
~~~~\\
\noindent
We discuss symmetry breaking in two-dimensional quantum dots resulting from
strong interelectron repulsion relative to the zero-point kinetic energy
associated with the confining potential. Such symmetry breaking leads to the
emergence of crystalline arrangements of electrons in the dot. The so-called
Wigner molecules form already at field-free conditions. The appearance of
rotating Wigner molecules in circular dots under high magnetic field, and their
relation to magic angular momenta and quantum-Hall-effect fractional fillings is
also discussed. Recent calculations for two electrons in an elliptic quantum dot,
using exact diagonalization and an approximate generalized-Heitler-London
treatment, show that the electrons can localize and form a molecular dimer for
screened interelectron repulsion. The calculated singlet-triplet splitting ($J$)
as a function of the magnetic field ($B$) agrees with cotunneling measurements; 
its behavior reflects the effective dissociation of the dimer for large $B$.
Knowledge of the dot shape and of $J(B)$ allows determination of two measures of
entanglement (concurrence and von Neumann entropy for 
{\it indistinguishable\/} fermions), whose behavior correlates 
also with the dissociation of the dimer. The theoretical value for the 
concurrence at $B=0$ agrees with the experimental estimates.  
\end{abstract}
\maketitle                   




\renewcommand{\leftmark}
{C. Yannouleas and U. Landman: Wigner molecules}

\section{Symmetry breaking in quantum dots}

Two-dimensional (2D) quantum dots (QDs), created at semiconductor interfaces 
through the use of lithographic and gate-voltage techniques, with refined 
control of their size, shape, and number of electrons, are often referred to as
``artificial atoms'' \cite{kast,taru2,asho}. These systems which, with the use 
of applied magnetic fields, are expected to have future applications as 
nanoscale logic gates and switching devices, have been in recent years the 
subject of significant theoretical and experimental research efforts. As 
indicated above, certain analogies have been made between these man-made systems
and their natural counterparts, suggesting that the physics of electrons in the 
former is similar to that underlying the traditional description of natural 
atoms -- pertaining particularly to electronic shells and the Aufbau principle 
in atoms (where electrons are taken to be moving in a spherically averaged 
effective central mean-field potential). 

The above-mentioned analogy has been theoretically challenged recently 
\cite{yyl1,yyl2} on the basis of calculations that showed evidence for 
formation, under favorable conditions (that are readily achieved in the 
laboratory), of ``electron molecules,'' which are alternatively called Wigner 
molecules (WMs) after the physicist who predicted formation of electron crystals
in extended systems \cite{wign}. These 
spin-and-space (sS) unrestricted Hartree-Fock (UHF) calculations (denoted in the
following as sS-UHF or simply UHF) of electrons confined in 2D QDs by a 
parabolic external potential led to the discovery of spontaneous symmetry 
breaking in QDs, manifested in the appearance of distinct interelectronic 
spatial (crystalline) correlations (even in the absence of magnetic fields). 
Such symmetry breaking may indeed be expected to occur based on the interplay 
between the interelectron repulsion, $Q$, and the zero-point kinetic energy, 
$K$. It is customary to take $Q=e^2/\kappa l_0$ and $K \equiv \hbar \omega_0$, 
where $l_0=(\hbar/m^*\omega_0)^{1/2}$ is the spatial extent of an electron in the
lowest state of the parabolic confinement; $m^*$ is the electron effective mass,
$\kappa$ is the dielectric constant, and $\omega_0$ is the frequency that 
characterizes the parabolic (harmonic) confining potential. Thus, defining the 
Wigner parameter as $R_W = Q/K$, one may expect symmetry breaking to occur when 
the interelectron repulsion dominates, i.e., for $R_W > 1$. Under such 
circumstances, an appropriate solution of the Schr\"{o}dinger equation 
necessitates consideration of wave functions with symmetries that are lower than
that of the circularly symmetric QD Hamiltonian. Such solutions may be found 
through the use of the sS-UHF method, where all restrictions on the symmetries 
of the wave functions are lifted. On the other hand, when $R_W < 1$, no symmetry
breaking is expected, and the sS-UHF solution collapses onto that obtained via 
the restricted Hartree-Fock (RHF) method, and the aforementioned circularly 
symmetric ``artificial-atom'' analogy maintains. From the above we note that the
state of the system may be controlled and varied through the choice of materials
(i.e., $\kappa$) and/or the strength of the confinement ($\omega_0$), since  
$R_W \propto 1/(\kappa \sqrt{\omega_0})$. 

\begin{figure}[t]
\centering\includegraphics[width=12cm]{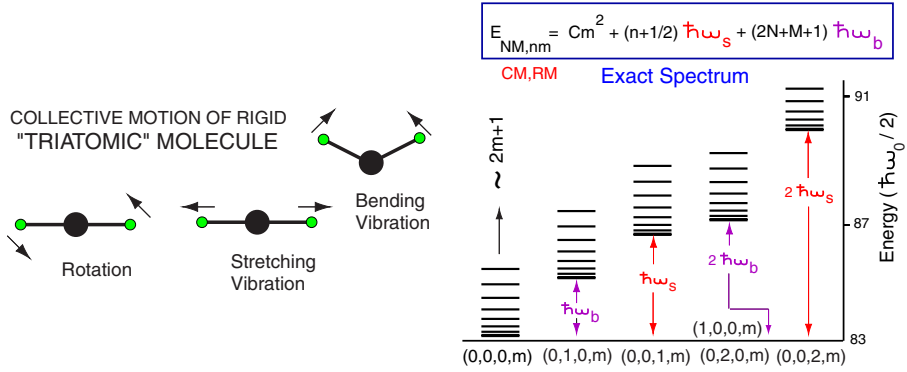}
\caption{The calculated spectrum of a two-electron parabolic quantum dot, with 
$R_W=200$. The quantum numbers are $(N, M, n, m)$ with $N$ corresponding to the 
number of radial nodes in the center of mass (CM) wavefunction, and $M$ is the 
CM azimuthal quantum number. The integers $n$ and $m$ are the corresponding 
quantum numbers for the electrons' relative motion (RM) and the total energy is 
given by $E_{NM,nm} = E^{cm}_{NM} + E^{rm}(n,|m|)$. The spectrum may be 
summarized by the ``spectral rule'' given in the figure, with $C = 0.037$, the 
phonon stretching vibration $\hbar \omega_s = 3.50$, and the phonon for the 
bending vibration coincides with that of the CM motion, i.e., $\hbar \omega_b 
= \hbar \omega_0 = 2$. All energies are in units of $\hbar \omega_0/2$, where 
$\omega_0$ is the parabolic confinement frequency.
}
\end{figure}

\subsection{Two-electron circular dots}

To illustrate the formation of ``electron molecules,'' we show first exact 
results obtained for a two-electron QD, through separation of the center-of-mass
and inter-electron relative-distance degrees of freedom \cite{yyl3}. The 
spectrum obtained for $R_W = 200$ (Fig.\ 1), exhibits features that are 
characteristic of a collective rovibrational dynamics, akin to that of a natural
``rigid'' triatomic molecule with an infinitely heavy middle particle 
representing the center of mass of the dot. This spectrum transforms to that of 
a ``floppy'' molecule for smaller value of $R_W$ (i.e., for stronger 
confinements characterized by a larger value of $\omega_0$, and/or for weaker 
inter-electron repulsion), ultimately converging to the independent-particle 
picture associated with the 
circular central mean-field of the QD. Further evidence for the formation of the
electron molecule was found through examination of the conditional probability 
distribution (CPD); that is, the anisotropic pair correlation 
$P({\bf r}, {\bf r}_0)$, which expresses the probability of finding a particle 
at ${\bf r}$ given that the ``observer'' (reference point) is riding on another 
particle at ${\bf r}_0$ \cite{yyl2,yyl3},
\begin{equation}
P({\bf r},{\bf r}_0) =
\langle \Psi | 
\sum_{i=1}^N \sum_{j \neq i}^N  \delta({\bf r}_i -{\bf r})
\delta({\bf r}_j-{\bf r}_0) 
| \Psi \rangle  / \langle \Psi | \Psi \rangle.
\label{cpds}
\end{equation} 
Here $\Psi ({\bf r}_1, {\bf r}_2, ..., {\bf r}_N)$ 
denotes the many-body wave function under consideration.
It is instructive to note here certain 
similarities between the formation of a ``two-electron molecule'' in man-made 
quantum dots, and the collective (rovibrational) features observed in the 
electronic spectrum of doubly-excited helium atoms \cite{kell1,kell2,berr}.

For confined (finite) systems with a larger number of particles, one must resort
to approximate computational schemes. Of particular interest are methodologies 
that permit systematic evaluation of high-accuracy solutions to these many-body 
strongly-correlated systems, under field-free conditions, as well as under the 
influence of an applied magnetic field. We remark here, that the relatively 
large (spatial) size of QDs (resulting from materials' characteristics, e.g., a 
small electron effective mass and large dielectric constant), allows the full 
range of orbital magnetic effects to be covered for magnetic fields that are 
readily attained in the laboratory (less then 40 T). In contrast, for natural 
atoms and molecules, magnetic fields of sufficient strength (i.e., larger than 
$10^5$ T) to produce novel phenomena related to orbital magnetism (beyond the 
perturbative regime), are known to occur only in astrophysical environments 
(e.g., on the surface of neutron stars).

The 2D hamiltonian of the problem under consideration is given by 
\begin{equation}
H=\sum_{i=1}^N \frac{1}{2m^*} \left( {\bf p}_i -\frac{e}{c} {\bf A}_i \right)^2
+ \sum_{i=1}^N \frac{m^*}{2} \omega_0^2 {\bf r}_i^2 +
\sum_{i=1}^N \sum_{j>i}^N \frac{e^2}{\kappa r_{ij}},
\label{mbhn} 
\end{equation}
describing $N$ electrons (interacting via a Coulomb repulsion) confined by a 
parabolic potential of frequency $\omega_0$ and subjected to a perpendicular 
magnetic field $ {\bf B}$, whose vector potential is given in the symmetric 
gauge by ${\bf A} ({\bf r}) = {\bf B} \times {\bf r}/2 = (-By,Bx,0)/2$.
For sufficiently high magnetic field values (i.e., in the fractional 
quantum Hall effect, or FQHE, regime), the electrons are fully spin-polarized 
and the Zeeman term (not shown here) does not need to be considered. 
In the $B \rightarrow \infty$ limit, the external confinement $\omega_0$ can be 
neglected, and $H$ can be restricted to operate in the lowest Landau level 
(LLL), reducing to the form \cite{yyl4,yyl5,yyl6,yyl7}
\begin{equation}
H_{\rm LLL} = N \frac{\hbar \omega_c}{2} +
\sum_{i=1}^N \sum_{j>i}^N \frac{e^2}{\kappa r_{ij}},
\label{hlll}
\end{equation}
where $\omega_c = eB/(m^*c)$ is the cyclotron frequency. 

For finite $N$, the solutions to the Schr\"{o}dinger equations corresponding to 
the hamiltonians given by Eq\. (\ref{mbhn}) (with or without a magnetic field), 
or by Eq\. (\ref{hlll}) (in the $B \rightarrow \infty$ limit), must have a good 
angular momentum, $L$, and good spin quantum numbers (the latter is guaranteed 
in the high $B$, fully spin-polarized case). As described in detail elsewhere 
\cite{yyl4,yyl5,yyl6,yyl7,yyl8,yyl9}, 
these solutions can be well approximated by a two-step method 
consisting of symmetry breaking at the spin-and-space unrestricted Hartree-Fock 
level and subsequent symmetry restoration via post-Hartree-Fock projection 
techniques. We recall that the sS-UHF method relaxes both the double-occupancy 
requirement (namely, different spatial orbitals are employed for different spin 
directions), as well as relaxing the requirement that the electron (spatial) 
orbitals be constrained by the symmetry of the external confining potential.

\begin{figure}[t]
\centering
\includegraphics[width=14.5cm]{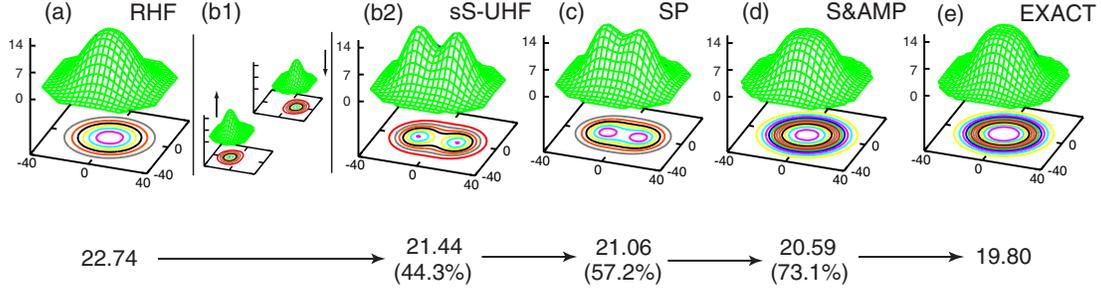}\\
~~~~~\\
\caption{Various approximation levels for a field-free two-electron QD with 
$R_W=2.40$. (a) Electron density of the RHF solution, exhibiting circular 
symmetry (due to the imposed symmetry restriction). The correlation energy 
$\varepsilon_c = 2.94$ meV, is defined as the difference between the energy of 
this state and the exact solution [shown in frame (e)]. (b1) and (b2) The two 
occupied orbitals (modulus square) of the symmetry-broken ``singlet'' sS-UHF 
solution (b1), with the corresponding total electron density exhibiting 
non-circular shape (b2). The energy of the sS-UHF solution shows a gain of 
44.3\% of the correlation energy. (c) Electron density of the spin-projected 
(SP) singlet, showing broken spatial symmetry, but with an additional gain of 
correlation energy. (d) the spin-and-angular-momentum projected state (S\&AMP) 
exhibiting restored circular symmetry with a 73.1\% gain of the correlation 
energy. The choice of parameters is: dielectric constant $\kappa = 8$, parabolic
confinement $\omega_0 = 5$ meV, and effective mass $m^* = 0.067m_e$. 
Distances are in nanometers and the densities in $10^{-4}$ nm$^{-2}$.
}
\end{figure}

Results obtained for various approximation levels for a two-electron QD with 
$B=0$ and $R_W=2.40$ (that is, in the Wigner-molecule regime) are displayed in 
Fig.\ 2. In these calculations \cite{yyl9}, the spin projection was performed 
following reference \cite{lowd}, i.e., one constructs the wave function
\begin{equation}
\Psi_{\rm SP}(s)={\cal P}_{\rm spin}(s) \Psi_{\rm UHF},
\label{psis}
\end{equation}
where $\Psi_{\rm UHF}$ is the original symmetry-broken UHF determinant.
In Eq. (\ref{psis}), the spin projection operator is given by
\begin{equation}
{\cal P}_{\rm spin}(s) \equiv \prod_{s^\prime \neq s}
\frac{\hat{S}^2 - s^\prime(s^\prime + 1) \hbar^2}
{[s(s+1) - s^\prime(s^\prime + 1)] \hbar^2}~,
\label{prjp}
\end{equation}
where the index $s^\prime$ runs over the quantum numbers of $\hat{S}^2$,
with $\hat{S}$ being the total spin.

The angular momentum projector is given by
\begin{equation}  
2\pi {\cal P}_L \equiv \int_0^{2\pi} d\gamma \exp[-i\gamma (\hat{L}-L)],
\label{amp}
\end{equation} 
where $\hat{L}=\hat{l}_1+\hat{l}_2$ is the total angular momentum 
operator. As seen from Eq.\ (\ref{amp}), application of the projection
operator ${\cal P}_L$ to the spin-restored state $\Psi_{\rm SP}(s)$
corresponds to a continuous configuration interaction (CCI) formalism. 

In the following we focus on the ground state of the system, i.e., 
$L = 0$. The energy of the projected state is given by 
\begin{equation}
E_{\rm PRJ}(L) = \left. \int_0^{2\pi} h(\gamma) e^{i\gamma L} d\gamma \right/ 
 \int_0^{2\pi} n(\gamma) e^{i\gamma L}d\gamma,
\label{eprj}
\end{equation}
with $h(\gamma)=\langle \Psi_{\rm SP}(s;0)|H|\Psi_{\rm SP}(s;\gamma)\rangle$
and $n(\gamma)=\langle \Psi_{\rm SP}(s;0)|\Psi_{\rm SP}(s;\gamma)\rangle$,
where $\Psi_{\rm SP}(s;\gamma)$ is the spin-restored wave function 
rotated by an azimuthal angle $\gamma$ and $H$ is the many-body hamiltonian.
We note that the UHF energies are simply given by $E_{\rm UHF}=h(0)/n(0)$.

The electron densities corresponding to the initial RHF approximation [shown in 
Fig.\ 2(a)] and the final spin-and-angular-momentum projection (S\&AMP) [shown 
in Fig.\ 2(d)], are circularly symmetric, while those corresponding to the two 
intermediate approximations, i.e., the sS-UHF and spin-projected (SP) solutions 
[Figs.\ 2(b2) and 2(c), respectively] break the circular symmetry. This 
behavior illustrates graphically the meaning of the term restoration of 
symmetry, and the interpretation that the sS-UHF broken-symmetry solution refers
to the intrinsic (rotating) frame of reference of the electron molecule. In 
light of this discussion the final projected state is called a {\it rotating 
Wigner molecule}, or RWM.

\subsection{Electrons in circular quantum dots under high magnetic fields}

To illustrate the emergence of RWMs in parabolically confined QDs under high $B$,
we show in Fig.\ 3 results obtained \cite{yyl7} 
through the aforementioned two-step 
computational technique for $N=7$, 8, and 9 electrons, and compare them with the
results derived from exact diagonalization of the Hamiltonian [see Eq\. 
(\ref{mbhn})]. Systematic investigations of QDs under high $B$ revealed 
electronic states of crystalline character. These states are found for 
particular ``magic'' angular momentum values ($L$) that exhibit enhanced 
stability and are called cusp states. For a given value of $B$, one of these 
$L$'s corresponds to the global minimum, i.e., the ground state, and varying $B$ 
causes the ground state and its angular momentum to change. The cusp states have
been long recognized \cite{laug1} as the finite-$N$ precursors of the fractional 
quantum Hall states in extended systems. In particular, the fractional fillings 
$\nu$ (defined in the thermodynamic limit) are related to the magic angular 
momenta of the finite-$N$ system as follows \cite{girv}
\begin{equation}
\nu=\frac{N(N-1)}{2L}.
\label{nu}
\end{equation}

In the literature of the fractional quantum Hall effect (FQHE), ever since the 
celebrated paper \cite{laug2} by Laughlin in 1983, the cusp states have been 
considered to be the antithesis of the Wigner crystal and to be described 
accurately by liquid-like wave functions, such as the Jastrow- Laughlin (JL) 
\cite{laug2,laug3} and composite-fermion (CF) \cite{jain,jain2} ones. This view,
however, has been challenged recently \cite{yyl4,yyl5} by the explicit 
derivation of trial wave functions for the cusp states that are associated with 
a rotating Wigner molecule. As discussed elsewhere \cite{yyl4,yyl5}, these 
parameter-free wave functions, which are by construction crystalline in 
character, have been shown to provide a simpler and improved description of the 
cusp states, in particular for high angular momenta (corresponding to low 
fractional fillings).

The crystalline arrangements that were found \cite{yyl4,yyl5,yyl6,yyl7} consist 
of concentric polygonal rings [see the conditional probabilities displayed in 
Fig.\ 3, with the reference (observation) point denoted by a filled dot]. These 
rings rotate independently of each other (see in particular the two cases shown 
for $N=9$), with the electrons on each ring rotating coherently \cite{yyl7}. The
rotations stabilize the RWM relative to the static one -- namely, the projected 
(symmetry restored) states are lower in energy compared to the broken-symmetry 
ones (the unrestricted Hartree-Fock solutions).

\begin{figure}[t]
\centering
\includegraphics[width=12.0cm]{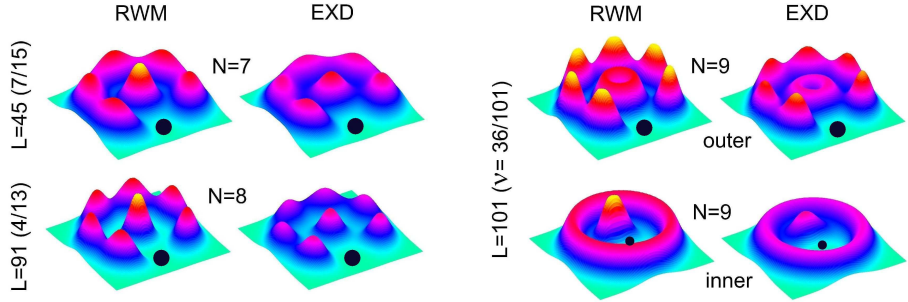}
\caption{Conditional probability distributions (CPDs) at high $B$, evaluated 
for parabolic quantum dots through: (i) the two-step procedure of symmetry 
breaking and subsequent restoration, resulting in rotating Wigner molecules 
(RWM) (shown in the left columns for given $N$), and (ii) exact diagonalization 
(EXD, shown in the right columns for given $N$). 
The angular momentum values, and corresponding values of the 
fractional filling [see Eq\. (\ref{nu})], are given on the left. The optimal 
polygonal structure for a given $N$ is given by $(n_1,n_2)$ with $n_1+n_2=N$. 
For $N= 7$, 8, and 9, these arrangements are (1,6), (1,7), and (2,7), 
respectively. The reference point for the calculation of the CPD is denoted by a
filled dot. Note in particular the two CPDs shown for $N=9$, illustrating that 
for a reference point located on the outer ring, the inner ring appears uniform,
and vice versa for a reference point located on the inner ring (bottom row on
the right). These results illustrate that the rings rotate independently of each
other. 
}
\end{figure}

\subsection{Symmetry breaking of trapped bosons}

In closing this section, we remark that the emergence of crystalline geometric 
arrangements, discussed above for electrons confined in 2D quantum dots, appears
to be a general phenomenon that is predicted \cite{roma} to occur also for 
trapped bosonic atomic systems (neutral or charged) when the interatomic 
repulsion is tuned to exceed the characteristic energy of the harmonic trap. 
Indeed, application of the aforementioned two-step method, allowed the 
evaluation \cite{roma} of solutions to the many-body Hamiltonians describing 
such bosonic systems, going beyond the mean-field (Gross-Pitaevskii) approach. 
These highly-correlated trapped 2D states exhibit localization of the bosons 
into polygonal-ringlike crystalline patterns, thus extending earlier work that 
predicted localization of strongly repelling 1D bosons (often referred to as the
Tonk-Girardeau regime \cite{gira1,gira2}), to a higher dimension. 
It is expected that these theoretical 
findings will be the subject of experimental explorations, using methodologies 
that have led recently to observations of localization transitions in 1D boson 
systems \cite{pare,weis}.

\section{Localization and entanglement in a two-electron elliptic 
quantum dot}

As discussed in the previous section, electron localization leading to formation
of molecular-like structures [the aforementioned Wigner molecules] within a 
{\it single circular\/} two-dimensional (2D) quantum dot at zero magnetic field 
($B$) has been theoretically predicted to occur \cite{yyl1,yyl2,yyl3,yl1,mikh}, 
as the strength of the $e$-$e$ repulsive interaction relative to the zero-point 
energy increases. Of particular interest is a two-electron $(2e)$ WM, in light 
of the current experimental effort \cite{taru,eng} aiming at implementation of a
spin-based \cite{burk} solid-state quantum logic gate that employs two coupled 
one-electron QDs (double dot). 

Here, we present an exact diagonalization (EXD) and an approximate
(generalized Heitler-London, GHL) microscopic treatment for two electrons in a
{\it single\/} elliptic QD specified by the 
parameters of a recently investigated experimental device \cite{marc}. While 
formation of Wigner molecules in circular QDs requires weak confinement, 
and thus large dots of lower densities (so that the interelectron repulsion 
dominates), we show that formation of such WMs is markedly enhanced in highly 
deformed (e.g., elliptic) dots due to their lower symmetry.
The calculations provide a good description
of the measured $J(B)$ curve (the singlet-triplet splitting) when screening 
\cite{kou2,hall} due to the metal gates and leads is included (in addition to 
the weakening of the effective inter-electron repulsion due to the
dielectric constant of the semiconductor, GaAs). In particular, our results
reproduce the salient experimental findings pertaining to the vanishing of 
$J(B)$ for a finite value of $B \sim 1.3 $ T [associated with a change in sign 
of $J(B)$ indicating a singlet-triplet (ST) transition], 
as well as the flattening of the $J(B)$ curve after the ST crossing.
These properties, and in particular the latter one, are related 
directly to the formation of an electron molecular dimer and its effective 
dissociation for large magnetic fields. 
The effective dissociation of the electron dimer is most naturally described 
through the GHL approximation, and it is fully supported by the more accurate,
but physically less transparent, EXD.

Of special interest for quantum computing is the degree of entanglement 
exhibited by the two-electron molecule in its singlet state \cite{burk}. 
Here, in relation to the microscopic calculations, we investigate two different 
measures of entanglement \cite{note27}. The first, known as the concurrence 
(${\cal C}$) for two {\it indistinguishable\/} fermions \cite{schl,loss}, 
has been used in the analysis of the experiment in Ref.\ \cite{marc}
(this measure is related to the operational cycle of a two-spin-qubit
quantum logic gate \cite{schl,loss}).
The second measure, referred to as the von Neumann entropy (${\cal S}$) for 
{\it indistinguishable\/} particles, has been developed in Ref.\ \cite{you}
and used in Ref.\ \cite{zung}.
We show that the present {\it wave-function-based\/} methods, in conjunction with
the knowledge of the dot shape and the $J(B)$ curve, enable theoretical 
determination of the degree of entanglement, in particular for the elliptic QD 
of Ref.\ \cite{marc}. The increase in the degree of entanglement (for both
measures) with stronger magnetic fields correlates with the dissociation
of the $2e$ molecule. This supports the experimental assertion \cite{marc} that 
cotunneling spectroscopy can probe properties of the electronic wave function of
the QD, and not merely its low-energy spectrum. Our methodology can be 
straightforwardly applied to other cases of strongly-interacting devices, e.g., 
double dots with strong interdot-tunneling. 

\subsection{Microscopic treatment}

The Hamiltonian for two 2D interacting electrons is [see Eq.\ (\ref{mbhn})]
\begin{equation}
{\cal H} = H({\bf r}_1)+H({\bf r}_2)+e^2/(\kappa r_{12}),
\label{ham}
\end{equation}
where the last term is the Coulomb repulsion, $\kappa$ is the
dielectric constant, and 
$r_{12} = |{\bf r}_1 - {\bf r}_2|$. $H({\bf r})$ is the
single-particle Hamiltonian for an electron in an external perpendicular
magnetic field ${\bf B}$ and an appropriate confinement potential.
When position-dependent screening is included, the last term in Eq.\
(\ref{ham}) is modified by a function of $r_{12}$ (see below). 
For an elliptic QD, the single-particle Hamiltonian is written as
\begin{equation}
H({\bf r}) = T + \frac{1}{2} m^* (\omega^2_{x} x^2 + \omega^2_{y} y^2)
    + \frac{g^* \mu_B}{\hbar} {\bf B \cdot s},
\label{hsp}
\end{equation}
where $T=({\bf p}-e{\bf A}/c)^2/2m^*$, with ${\bf A}=0.5(-By,Bx,0)$ being the
vector potential in the symmetric gauge. $m^*$ is the effective mass and
${\bf p}$ is the linear momentum of the electron. The second term is the
external confining potential; the last term is the Zeeman interaction with 
$g^*$ being the effective $g$ factor, $\mu_B$ the Bohr magneton, and ${\bf s}$ 
the spin of an individual electron. 

The GHL method for solving the Hamiltoninian (\ref{ham}) consists of two steps. 
In the first step, we solve selfconsistently the ensuing 
unrestricted Hartree-Fock (UHF) equations allowing for lifting of the 
double-occupancy requirement (imposing this requirement gives the 
{\it restricted\/} HF method, RHF).
For the $S_z=0$ solution, this step produces two single-electron 
orbitals $u_{L,R}({\bf r})$ that are localized left $(L)$ and right $(R)$ of the
center of the QD [unlike the RHF method that gives a single doubly-occupied 
elliptic (and symmetric about the origin) orbital]. 
At this step, the many-body wave function is a single Slater 
determinant $\Psi_{\text{UHF}} (1\uparrow,2\downarrow) \equiv 
| u_L(1\uparrow)u_R(2\downarrow) \rangle$ made out of the two occupied UHF 
spin-orbitals $u_L(1\uparrow) \equiv u_L({\bf r}_1)\alpha(1)$ and 
$u_R(2\downarrow) \equiv u_R({\bf r}_2) \beta(2)$, where 
$\alpha (\beta)$ denotes the up (down) [$\uparrow (\downarrow)$] spin. 
This UHF determinant is an eigenfunction of the projection $S_z$ of the total 
spin $\hat{S} = \hat{s}_1 + \hat{s}_2$, but not of $\hat{S}^2$ (or the parity
space-reflection operator). 

In the second step, we restore the broken parity and total-spin symmetries by 
applying to the UHF determinant the projection operator \cite{yyl8,yl3} 
${\cal P}_{\rm spin}^{s,t}=1 \mp \varpi_{12}$, where the 
operator $\varpi_{12}$ interchanges the spins of the two electrons
[this is a special case of the operator given in Eq.\ (\ref{prjp})]; 
the upper (minus) sign corresponds to the singlet. 
The final result is a generalized Heitler-London (GHL) two-electron wave function
$\Psi^{s,t}_{\text{GHL}} ({\bf r}_1, {\bf r}_2)$ for the ground-state singlet 
(index $s$) and first-excited triplet (index $t$), which uses
the UHF localized orbitals,
\begin{equation}
\Psi^{s,t}_{\text{GHL}} ({\bf r}_1, {\bf r}_2) \propto
{\bf (} u_L({\bf r}_1) u_R({\bf r}_2) \pm u_L({\bf r}_2) u_R({\bf r}_1) {\bf )}
\chi^{s,t},
\label{wfghl}
\end{equation}
where $\chi^{s,t} = (\alpha(1) \beta(2) \mp \alpha(2) \beta(1))$ is the spin 
function for the 2$e$ singlet and triplet states.
The general formalism of the 2D UHF equations and of the subsequent restoration 
of broken spin symmetries can be found in Refs.\ \cite{yyl8,yyl9,yl1,yl3}.

The use of {\it optimized\/} UHF orbitals in the GHL is suitable for treating 
{\it single elongated\/} QDs. The GHL is equally applicable to double QDs with 
arbitrary interdot-tunneling coupling \cite{yyl8,yl3}. In contrast,
the Heitler-London (HL) treatment \cite{hl} (known also as Valence bond), 
where non-optimized ``atomic'' orbitals of two isolated QDs are used, is 
appropriate only for the case of a double dot with small interdot-tunneling 
coupling \cite{burk}.

The orbitals $u_{L,R}({\bf r})$ are expanded in a real Cartesian 
harmonic-oscillator basis, i.e.,
\begin{equation}
u_{L,R}({\bf r}) = \sum_{j=1}^K C_j^{L,R} \varphi_j ({\bf r}),
\label{uexp}
\end{equation}
where the index $j \equiv (m,n)$ and $\varphi_j ({\bf r}) = X_m(x) Y_n(y)$,
with $X_m(Y_n)$ being the eigenfunctions of the one-dimensional oscillator in the
$x$($y$) direction with frequency $\omega_x$($\omega_y$). The parity operator
${\cal P}$ yields ${\cal P} X_m(x) = (-1)^m X_m(x)$, and similarly for $Y_n(y)$.
The expansion coefficients $C_j^{L,R}$ are real for $B=0$ and complex for finite
$B$. In the calculations we use $K=79$, yielding convergent results.

In the EXD method, the many-body wave function is written as a linear
superposition over the basis of non-interacting two-electron determinants, i.e.,
\begin{equation}
\Psi^{s,t}_{\text{EXD}} ({\bf r}_1, {\bf r}_2) =
\sum_{i < j}^{2K} \Omega_{ij}^{s,t} | \psi(1;i) \psi(2;j)\rangle,
\label{wfexd}
\end{equation}
where $\psi(1;i) = \varphi_i(1 \uparrow)$ if $1 \leq i \leq K$ and
$\psi(1;i) = \varphi_{i-K}(1 \downarrow)$ if $K+1 \leq i \leq 2K$ [and
similarly for $\psi(2,j)$].
The total energies $E^{s,t}_{\text{EXD}}$ and the coefficients
$\Omega_{ij}^{s,t}$ are obtained through a ``brute force'' diagonalization of
the matrix eigenvalue equation corresponding to the Hamiltonian in Eq.\ 
(\ref{ham}). The EXD wave function does not
immediately reveal any particular form, although, our calculations below
show that it can be approximated by a GHL wave function in the case of the
elliptic dot under consideration. 

\begin{figure}[t]
\centering
\includegraphics[width=6.5cm]{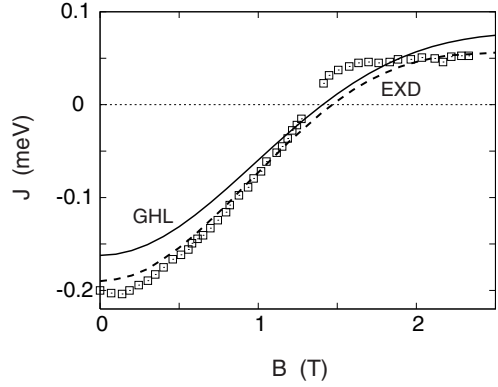}    
\caption{
The singlet-triplet splitting $J=E^s-E^t$ as a function of the magnetic field
$B$ for an elliptic QD with $\hbar \omega_x=1.2$ meV and $\hbar \omega_y=3.3$ meV
(these values correspond to the device of Ref.\ \cite{marc}).
Solid line: GHL (broken-symmetry UHF + restoration of symmetries) results
with a coordinate-independent screening ($\kappa=22$).
Dashed line: EXD results with $\kappa=12.9$ (GaAs), but including screening with
a coordinate dependence according to Ref.\ \cite{hall} and $d=18.0$ nm 
(see text). The rest of the material parameters used are: 
$m^*$(GaAs)$=0.067 m_e$, and $g^*=0$ (see text).
The experimental measurements \cite{marc} are denoted by open squares.
Our sign convention for $J$ is opposite to that in Ref.\ \cite{marc}.
}
\end{figure}

To model the experimental elliptic QD device, we take, following Ref.\ 
\cite{marc}, $\hbar \omega_x=1.2$ meV and $\hbar \omega_y=3.3$ meV. 
The effective mass of the electron is taken as $m^*=0.067 m_e$ (GaAs). Since 
the experiment did not resolve the lifting of the triplet degeneracy caused by 
the Zeeman term, we take $g^*=0$. Using the two step method, 
we calculate the GHL singlet-triplet splitting 
$J_{\text{GHL}}(B)=E^s_{\text{GHL}}(B)-E^t_{\text{GHL}}(B)$ 
as a function of the magnetic field in the range $0 \leq B \leq 2.5$ T.
Screening of the $e$-$e$ interaction due to the metal gates and leads 
must be considered in order to reproduce the experimental $J(B)$ 
curve \cite{note2}. This screening can be modeled, to first approximation, by a 
position-independent adjustment of the dielectric constant 
$\kappa$ \cite{kyri}. Indeed, with $\kappa=22.0$ (instead of the GaAs 
dielectric constant, i.e., $\kappa = 12.9$), good agreement with
the experimental data is obtained [see Fig.\ 4]. In particular, we note the 
singlet-triplet crossing for $B \approx 1.3$ T, and the flattening of the 
$J(B)$ curve beyond this crossing.

We have also explored, particularly in the context of the EXD treatment, 
a position-dependent screening using the functional form,
$(e^2/\kappa r_{12}) [1-(1+4d^2/r_{12}^2)^{-1/2}]$, 
proposed in Ref.\ \cite{hall}, with $d$ as a fitting parameter. 
The $J_{\text{EXD}}(B)$ result for $d=18.0$ nm is depicted in Fig.\ 4 
(dashed line), and it is in very good agreement with the experimental 
measurement.

\begin{figure}[t]
\centering
\includegraphics[width=6.5cm]{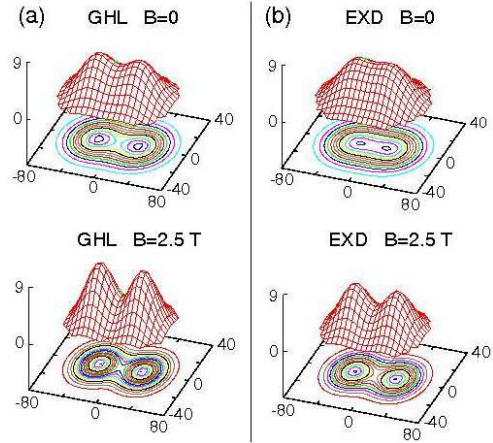}    
\caption{ 
Total electron densities (EDs) associated with the singlet state of the
elliptic dot at $B=0$ and $B=2.5$ T.
(a) The GHL densities. (b) The EXD densities.
The rest of the parameters and the screening of the Coulomb interaction 
are as in Fig.\ 4.
Lengths in nm and densities in $10^{-4}$ nm$^{-2}$.
}
\end{figure}


The singlet state electron densities from the GHL and the EXD treatments
at $B=0$ and $B=2.5$ T are displayed in Fig.\ 5. These densities illustrate the 
dissociation of the electron dimer with increasing magnetic field. 
The asymptotic convergence (beyond the ST point) of the energies of the singlet 
and triplet states, i.e., [$J(B) \rightarrow 0$ as $B \rightarrow \infty$] is a 
reflection of the dissociation of the 2$e$ molecule, since the ground-state 
energy of two fully spatially separated electrons (zero overlap) does not 
depend on the total spin \cite{note67}. 

\subsection{Measures of entanglement \cite{note27}}

To calculate the concurrence ${\cal C}$ \cite{schl,loss}, one needs a 
decomposition of the GHL wave function into a 
linear superposition of {\it orthogonal\/}
Slater determinants. Thus one needs to expand the {\it nonorthogonal\/} 
$u^{L,R}({\bf r})$ orbitals as a superposition of two other {\it orthogonal\/}
ones. To this effect, we write
$u^{L,R}({\bf r}) \propto \Phi^+({\bf r}) \pm  \xi \Phi^-({\bf r})$,
where $\Phi^+({\bf r})$ and $\Phi^-({\bf r})$ are the parity symmetric and 
antisymmetric (along the $x$-axis) components in their expansion given by 
Eq.\ (\ref{uexp}). 
Subsequently, with the use of Eq.\ (\ref{wfghl}), the GHL 
singlet can be rearranged as follows:
\begin{equation}
\Psi^{s}_{\text{GHL}} \propto
| \Phi^+(1\uparrow) \Phi^+(2\downarrow) \rangle - 
\eta |\Phi^-(1\uparrow)\Phi^-(2\downarrow) \rangle,
\label{rear}
\end{equation}
where the so-called interaction parameter \cite{loss}, $\eta=\xi^2$, is the 
coefficient in front of the second determinant.
Knowing $\eta$ allows a direct evaluation of the concurrence of the singlet
state, since ${\cal C}^s = 2\eta/(1+\eta^2)$ \cite{loss}. Note that 
$\Phi^+({\bf r})$ and $\Phi^-({\bf r})$ are properly normalized.
It is straightforward to show that $\eta=(1-|S_{LR}|)/(1+|S_{LR}|)$, where 
$S_{LR}$ (with $|S_{LR}| \leq 1$) is the overlap of the original 
$u^{L,R}({\bf r})$ orbitals.

For the GHL triplet, one obtains an expression independent of the 
interaction parameter $\eta$, i.e.,
\begin{equation}
\Psi^{t}_{\text{GHL}} \propto
| \Phi^+(1\uparrow) \Phi^-(2\downarrow) \rangle +
|\Phi^+(1\downarrow)\Phi^-(2\uparrow) \rangle,
\label{reart}
\end{equation}
which is a maximally (${\cal C}^t=1$) entangled state. Note that underlying the 
analysis of the experiments in Ref.\ \cite{marc} is a {\it conjecture\/} that 
wave functions of the form given in Eqs.\ (\ref{rear}) and (\ref{reart}) 
describe the two electrons in the elliptic QD.

For the GHL singlet, using the overlaps of the left and right orbitals, we find
that starting with $\eta=0.46$ $({\cal C}^s=0.76)$ at $B=0$, the interaction
parameter (singlet-state concurrence) increases monotonically to $\eta=0.65$ 
$({\cal C}^s=0.92)$ at $B=2.5$ T. At the intermediate value corresponding to the
ST transition ($B=1.3$ T), we find $\eta=0.54$ $({\cal C}^s=0.83)$ 
\cite{note34}. Our $B=0$ theoretical results for 
$\eta$ and ${\cal C}^s$ are in remarkable agreement with 
the experimental estimates \cite{marc} of $\eta=0.5 \pm 0.1$ and 
${\cal C}^s=0.8$, which were based solely on conductance measurements below the 
ST transition (i.e., near $B=0$). 

To compute the von Neumann entropy, one needs to bring both 
the EXD and the GHL wave functions into a diagonal form (the socalled 
``canonical form'' \cite{you,schl2}), i.e.,
\begin{equation}
\Psi^{s,t}_{\text{EXD}} ({\bf r}_1, {\bf r}_2) =
\sum_{k=1}^M z^{s,t}_k | \Phi(1;2k-1) \Phi(2;2k) \rangle,
\label{cano}
\end{equation}
with the $\Phi(i)$'s being appropriate spin orbitals resulting from a unitary 
transformation of the basis spin orbitals $\psi(j)$'s [see Eq.\ (\ref{wfexd})]; 
only terms with $z_k \neq 0$ contribute. The upper bound $M$ can be 
smaller (but not larger) than $K$ (the dimension of the 
single-particle basis); $M$ is referred to as the Slater rank.
One obtains the coefficients of the canonical expansion from the fact that 
the $|z_k|^2$ are eigenvalues of the hermitian matrix $\Omega^\dagger \Omega$
[$\Omega$, see Eq.\ (\ref{wfexd}), is antisymmetric]. The von Neumann 
entropy is given by ${\cal S} = -\sum_{k=1}^M |z_k|^2 \log_2(|z_k|^2)$ with the
normalization $\sum_{k=1}^M |z_k|^2 =1$.
Note that the GHL wave functions in Eqs.\ (\ref{rear}) and (\ref{reart}) 
are already in canonical form, which shows that they always have a Slater rank 
of $M=2$. One finds ${\cal S}^s_{\text{GHL}} =
\log_2(1+\eta^2) - \eta^2 \log_2(\eta^2)/(1+\eta^2)$, and
${\cal S}^t_{\text{GHL}}=1$ for all $B$. For large $B$, the overlap 
between the two electrons of the dissociated dimer vanishes, and thus 
$\eta \rightarrow 1$ and ${\cal S}^s_{\text{GHL}} \rightarrow 1$.

\begin{figure}[t]
\centering\includegraphics[width=10cm]{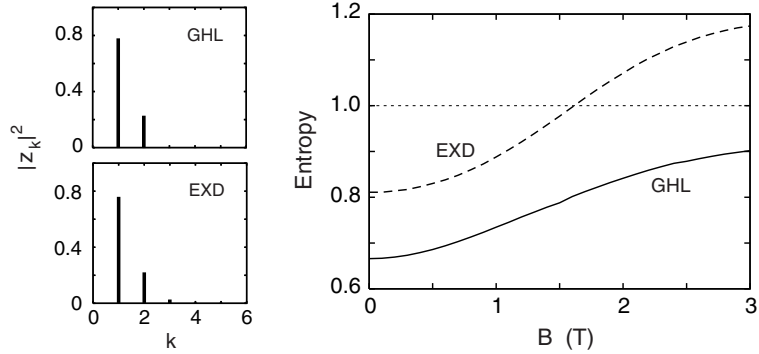}
\caption{
Von Neumann entropy for the singlet state of the elliptic dot as a function of 
the magnetic field $B$. Solid line: GHL. Dashed line: EXD. 
The rest of the parameters and the screening of the Coulomb interaction 
are as in Fig.\ 4. On the left, we show histograms for the $|z_k|^2$
coefficients [see Eq.\ (\ref{cano})] of the singlet state at $B=1.3$ T, 
illustrating the dominance of two configurations. Note the small third
coefficient $|z_3|^2=0.023$ in the EXD case.
}
\end{figure}

Since the EXD singlet has obviously a Slater rank $M > 2$, the definition
of concurrence is not applicable to it.  
The von Neumann entropy for the EXD singlet (${\cal S}^s_{\text{EXD}}$) is 
displayed in Fig.\ 6, along with that (${\cal S}^s_{\text{GHL}}$) of the GHL 
singlet. ${\cal S}^s_{\text{EXD}}$ and 
${\cal S}^s_{\text{GHL}}$ are rather close to each other for the entire
$B$ range, and it is remarkable that both 
remain close to unity for large $B$, although the maximum allowed mathematical 
value is $\log_2(K)$ [as aforementioned we use $K=79$, i.e., $\log_2(79)=6.3$];
this maximal value applies for both the EXD and GHL approaches. The saturation 
of the entropy for large $B$ to a value close to unity reflects the dominant 
(and roughly equal at large $B$) weight of two configurations in the 
canonical expansion [see Eq.\ (\ref{cano})] of the EXD wave function, which
are related to the two terms ($M=2$) in the canonical expansion of the GHL
singlet [Eq.\ (\ref{rear})]. This is illustrated by the histograms of the
$|z_k^s|^2$ coefficients for $B=1.3$ T in Fig.\ 6 (left column).
These observations support the GHL approximation, which is
computationally less demanding than the exact diagonalization, and can be used
easily for larger $N$.

\section{Summary}

We discussed symmetry breaking in two-dimensional quantum dots resulting from
strong interelectron repulsion relative to the zero-point kinetic energy
associated with the confining potential. Such symmetry breaking leads to the
emergence of crystalline arrangements of electrons in the dot. The so-called
Wigner molecules form already at field-free conditions. The appearance of
rotating Wigner molecules in circular dots under high magnetic field, and their
relation to magic angular momenta and quantum-Hall-effect fractional fillings was
also discussed. 

Furthermore, we have shown, through exact and approximate microscopic
treatments, formation of an electron molecular dimer in an
elliptic QD (Fig.\ 5) for screened interelectron repulsion.
The formation and effective dissociation (in high magnetic fields) of the 
electron dimer are reflected in the behavior of the computed singlet-triplet
splitting, $J(B)$, that agrees well (see Fig.\ 4) with the measurements 
\cite{marc}. Furthermore, we showed that, from a knowledge of the dot shape and 
of $J(B)$, theoretical analysis along the lines introduced here allows probing 
of the correlated ground-state wave function and determination of its degree of 
entanglement. This presents an alternative to the experimental study where
determination of the concurrence utilized conductance data \cite{marc}.
We have employed two measures of entanglement for {\it
indistinguishable fermions\/} (the concurrence and the von Neumann entropy) 
and have shown that their behavior correlates with the effective dissociation of
the electron dimer. Such information is of interest to the implementation of 
spin-based solid-state quantum logic gates.

This research is supported by the US D.O.E. (Grant No. FG05-86ER45234), and
NSF (Grant No. DMR-0205328). We thank M. Pustilnik for comments on the
manuscript.


\begin{thebibliography}{99}
\bibitem{kast}
M.A. Kastner,
Physics Today {\bf 46}, 24 (1993).
\bibitem{taru2}
S. Tarucha {\it et al.\/} 
Phys. Rev. Lett. {\bf 77}, 3613 (1996).
\bibitem{asho}
R.C. Ashoori,
Nature {\bf 379}, 413 (1996).
\bibitem{yyl1}
C. Yannouleas and U. Landman,
Phys. Rev. Lett. {\bf 82}, 5325 (1999);
{\bf 85}, E2220 (2000).
\bibitem{yyl2}
C. Yannouleas and U. Landman,
Phys. Rev. B {\bf 61}, 15895 (2000).
\bibitem{wign}
E. Wigner,
Phys. Rev. {\bf 46}, 1002 (1934).
\bibitem{yyl3}
C. Yannouleas and U. Landman,
Phys. Rev. Lett. {\bf 85}, 1726 (2000).
\bibitem{kell1}
M.E. Kellman and D.R. Herrick,
Phys. Rev. A {\bf 22}, 1536 (1980). 
\bibitem{kell2}
M.E. Kellman,
Int. J. Quantum Chem. {\bf 65}, 399 (1997).
\bibitem{berr}
R.S. Berry,
Contemp. Phys. {\bf 30}, 1 (1989).
\bibitem{yyl4}
C. Yannouleas and U. Landman,
Phys. Rev. B {\bf 66}, 115315 (2002).
\bibitem{yyl5}
C. Yannouleas and U. Landman,
Phys. Rev. B {\bf 68}, 035326 (2003).
\bibitem{yyl6}
C. Yannouleas and U. Landman,
Phys. Rev. B {\bf 69}, 113306 (2004).
\bibitem{yyl7}
C. Yannouleas and U. Landman,
Phys. Rev. B {\bf 70}, 235319 (2004).
\bibitem{yyl8}
C. Yannouleas and U. Landman,
Int. J. Quantum Chem. {\bf 90}, 699 (2002).
\bibitem{yyl9}
C. Yannouleas and U. Landman,
J. Phys.: Condens. Matter {\bf 14}, L591 (2002).
\bibitem{lowd}
P.O. L\"{o}wdin,
Rev. Mod. Phys. {\bf 36}, 966 (1964).
\bibitem{laug1}
R.B. Laughlin,
Phys. Rev. B {\bf 27}, 3383 (1983).
\bibitem{girv}
S.M. Girvin and T. Jach,
Phys. Rev. B {\bf 28}, 4506 (1983).
\bibitem{laug2}
R.B. Laughlin,
Phys. Rev. Lett. {\bf 50}, 1395 (1983).
\bibitem{laug3}
R.B. Laughlin,
in: The Quantum Hall Effect,edited by R.E. Prange and S.M. Girvin
(Springer, New York, 1987), p. 233.
\bibitem{jain}
J.K. Jain,
Phys. Rev. B {\bf 41}, 7653 (1990).
\bibitem{jain2}
S.S. Mandal, M.R. Peterson, and J.K. Jain,
Phys. Rev. Lett. {\bf 90}, 106403 (2003).
\bibitem{roma}
I. Romanovsky, C. Yannouleas, and U. Landman,
Phys. Rev. Lett. {\bf 93}, 230405 (2004).
\bibitem{gira1}
M. Girardeau,
J. Math. Phys. {\bf 1}, 516 (1960).
\bibitem{gira2}
M.D. Girardeau and E.M. Wright,
Laser Phys. {\bf 12}, 8 (2002).
\bibitem{pare}
B. Paredes {\it et al.\/},
Nature {\bf 429}, 277 (2004).
\bibitem{weis}
G.T. Kinoshita, T. Wenger, and D.S. Weiss,
Science {\bf 305}, 1125 (2004).
\bibitem{yl1}
C. Yannouleas and U. Landman,
Phys. Rev. B {\bf 68}, 035325 (2003).
\bibitem{mikh}
S.A. Mikhailov,
Phys. Rev. B {\bf 65}, 115312 (2002).
\bibitem{taru}
J.M. Elzerman {\it et al.\/},
Lect. Notes Phys. {\bf 667}, 25 (2005),
edited by W.D. Heiss (Springer, Berlin, 2005).
\bibitem{eng}
H.A. Engel {\it et al.\/},
Quantum Information Processing {\bf 3}, 115 (2004).
\bibitem{burk}
G. Burkard, D. Loss, and D.P. DiVincenzo,
Phys. Rev. B {\bf 59}, 2070 (1999).
\bibitem{marc}
D.M. Zumb\"{u}hl {\it et al.\/},
Phys. Rev. Lett. {\bf 93}, 256801 (2004). 
\bibitem{kou2}
S. Tarucha, D.G. Austing, Y. Tokura, W.G. van der Wiel, and L.P. Kouwenhoven,
Phys. Rev. Lett. {\bf 84}, 2485 (2000). 
\bibitem{hall}
L.D. Hallam, J. Weis, and P.A. Maksym,
Phys. Rev. B {\bf 53}, 1452 (1996).
\bibitem{note27}
Various approaches for estimating the degree of entanglement have
been proposed, including schemes which apply for {\it distinguishable\/}
[see e.g., W.K. Wootters, Phys. Rev. Lett. {\bf 80}, 2245 (1998)]
or {\it indistinguishable\/} particles, with only the latter being relevant in
our case. For a review, see K. Eckert {\it et al.\/}, Ann. of Phys.
(NY) {\bf 299}, 88 (2002); for the consequences of indistinguishability,
see section 2.2 therein.
\bibitem{schl}
J. Schliemann, D. Loss, and A.H. MacDonald,
Phys. Rev. B {\bf 63}, 085311 (2001).
\bibitem{loss}
V.N. Golovach and D. Loss,
Phys. Rev. B {\bf 69}, 245327 (2004).
\bibitem{you}
R. Paskauskas and L. You,
Phys. Rev. A {\bf 64}, 042310 (2001).
\bibitem{zung}
L. Xe, G. Bester, and A. Zunger,
cond-mat/0503492
\bibitem{yl3}
C. Yannouleas and U. Landman,
Eur. Phys. J. D {\bf 16}, 373 (2001).
\bibitem{hl}
H. Heitler and F. London,
Z. Phys. {\bf 44}, 455 (1927).
\bibitem{note2}
This is in agreement with the analysis of the only other measured $J(B)$ curve 
in a different device \cite{kyri}, where the $J(B)$ curve was measured 
{\it before\/} the singlet-triplet crossing.
\bibitem{kyri}
J. Kyriakidis {\it et al.\/},
Phys. Rev. B {\bf 66}, 035320 (2002)].
\bibitem{note67}
In contrast, the singlet-state RHF electron densities fail to exhibit 
formation of an electron dimer for all values of $B$. This underlies the
failure of the RHF method to describe the behavior of the 
experimental $J(B)$ curve. In particular, $J_{\text{RHF}}(B=0)$ has the
wrong sign, while $J_{\text{RHF}}(B)$ diverges for high $B$ as is the case
for the RHF treatment of double dots (see Ref.\ \cite{yyl8,yl3}).
\bibitem{note34}
For the RHF, ${\cal C}^s_{\text{RHF}}=0$, since a single determinant 
is unentangled for both the two measures considered here. 
\bibitem{schl2}
J. Schliemann {\it et al.\/},
Phys. Rev. A {\bf 64}, 022303 (2001). 
\end{thebibliography}
\end{document}